\documentclass{PoS}

\newcommand{\be}{\begin{equation}}
\newcommand{\ee}{\end{equation}}
\newcommand{\bea}{\begin{eqnarray}}
\newcommand{\eea}{\end{eqnarray}}
\newcommand{\beas}{\begin{eqnarray*}}
\newcommand{\eeas}{\end{eqnarray*}}

\def\({\left(}
\def\){\right)}

\newcommand{\rd}{{\rm d}}
\newcommand{\vp}{\varphi}

\def\gsim{ \lower .75ex \hbox{$\sim$} \llap{\raise .27ex \hbox{$>$}} }
\def\lsim{ \lower .75ex \hbox{$\sim$} \llap{\raise .27ex \hbox{$<$}} }

\title{Theoretical Aspects of Cosmic Acceleration}

\ShortTitle{Theoretical Aspects of Cosmic Acceleration}

\author{\speaker{Mark Trodden}%
        \\
       Center for Particle Cosmology, Department of Physics and Astronomy, \\ University of Pennsylvania, Philadelphia, PA 19104, USA.\\
       E-mail: \email{trodden@upenn.edu}}

\abstract{Efforts to understand and map the possible explanations for the late time acceleration of the universe have led to a broad range of suggestions, ranging from the cosmological constant and straightforward dark energy, to exotically coupled models, to infrared modifications of General Relativity. If we are to uncover which, if any, of these approaches might provide a serious answer to the problem, it is crucial to understand the constraints that theoretical consistency places on the models, and on the regimes in which they make predictions. In this talk, delivered as an invited plenary lecture at the Dark Side of the Universe conference in Kyoto, Japan, I briefly describe some modern attempts to carry out this program and some of the more interesting ideas that have emerged. As an example, I use the Galileon model, discussing how the Vainshtein mechanism occurs, and how a number of these theoretical problems arise around such backgrounds.}

\FullConference{11th International Workshop Dark Side of the Universe 2015\\
		 14-18 December 2015\\
		 Yukawa Institute for Theoretical Physics, Kyoto University Japan}

\begin{document}

\section{Introduction}
General Relativity (GR), applied to a universe populated by matter of relatively ordinary types, such as baryons, dark matter, photons, and neutrinos, can only yield expanding solutions that are decelerating. Any attempt to explain the currently observed accelerating expansion rate must therefore go beyond this simple paradigm, typically in one of three ways. Nominally the simplest, but still yet to be understood, idea is to include the cosmological constant (CC), which, if positive, allows for acceleration and agreement with all current data. While this is certainly a possibility, there are at present no known dynamical ways to generate a CC of the required magnitude. Anthropic arguments may be invoked as an alternative, but at this stage we do not know if these are viable, even within string theory, and considerably more work is required to understand this question. A second possible approach is to include some kind of exotic mass-energy component in the universe, with dynamics such that it can drive accelerated expansion. Such approaches go under the heading of {\it dark energy} and require an additional layer of complexity, since the problem of why the CC is essentially zero must also be solved (usually assumed) in a different way in these models.

A third approach is to leave the matter content of the universe essentially untouched, and instead to imagine that the physical laws connecting this to geometry are modified from those of GR at cosmological distances - in the far infrared. These approaches are typically referred to as {\it modified gravity} and have seen increased interest in recent years, in part because they contain at least the possibility that the CC problem might be addressed as part of the model that explains acceleration, possibly justifying the superficial complexity of these theories.

Whether dark energy or modified gravity proves to be on the right track, from the point of view of particle physics they face similar challenges. Around a given background solution (the accelerating universe, the early universe, spherically symmetric solutions, gravitational waves, ...) we expect that there should exist a sensible effective field theory (EFT) description of the relevant degrees of freedom, which is constrained by the rules of a well-behaved quantum field theory.

In this summary of a plenary lecture, delivered at the 2015 Dark Side of the Universe conference, I describe in broad terms, accompanied by some examples, how these considerations shape the construction of dark energy models. After describing a subset of the theoretical constraints, I focus on a particularly interesting class of models exhibiting {\it Vainshtein screening} - the {\it Galileons} and sketch how screening occurs and how some constraints result.

As is usual in a proceedings, I will have space only to reference those papers of most direct relevance to my own thinking on the subject, and will not attempt to be comprehensive. In particular, I will draw heavily in this text from the extensive review article~\cite{Joyce:2014kja} which I co-authored.

\section{Theoretical Consistency}
Here I detail some of the more common pathologies at the level of the effective field theory, namely ghosts, gradient instabilities, tachyons and superluminality.

\subsection{Ghosts}

Ghosts are fields whose quanta either have negative energy or negative norm, indicating an instability in the theory. Most commonly, ghost instabilities manifest as fields with {\it wrong sign} kinetic term
\be
{\cal L}_{\rm ghost} = \frac{1}{2}(\partial\chi)^2-\frac{m^2}{2}\chi^2~.
\ee
Of course, the sign of the kinetic term is merely a matter of convention---the choice of metric signature. However, what is dangerous is if this field is coupled to other fields with correct sign kinetic terms, for example
\be
{\cal L} = -\frac{1}{2}(\partial\phi)^2-\frac{m_\phi^2}{2}\phi^2 + \frac{1}{2}(\partial\chi)^2-\frac{m_\chi^2}{2}\chi^2+\lambda\phi^2\chi^2~.
\ee
Since the $\chi$ particles have negative energy, the vacuum is unstable to the process $0 \to \phi\phi+\chi\chi$, which costs zero energy. This process will happen copiously (with an infinite rate), and in fact is a sign that the theory is ill-defined~\cite{Carroll:2003st, Cline:2003gs}.

\begin{figure}
\centering
\includegraphics[width=2in]{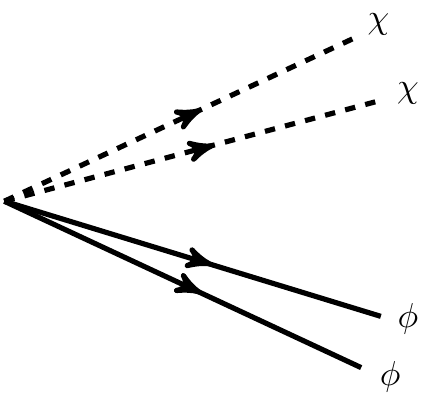}
\caption{\small In theories with a ghostly field, the vacuum is unstable to rapid pair production of ghost particles and healthy particles, causing the theory to be ill-defined (See~\cite{Carroll:2003st,Cline:2003gs}).}
\end{figure}

In constructing theories, we must ensure that the theory is free from these ghost instabilities. Most often, ghost instabilities arise from {\it higher derivative} terms in the Lagrangian. A powerful theorem (due to Ostrogradsky) then tells us that, in most cases, if the equations of motion are higher than second order in time derivatives, the theory will have a ghost instability.

To illustrate this, consider a simple example of a higher-derivative theory, with Lagrangian
\be
{\cal L} = -\frac{1}{2}(\partial\psi)^2+\frac{1}{2\Lambda^2}(\square\psi)^2-V(\psi)~,
\label{simpleghostlag}
\ee
where $\Lambda$ is the cutoff of the effective theory. As long as we work at energies far below $\Lambda$, this theory is completely well defined. However, as we approach $\Lambda$, all of the terms we have neglected---which are suppressed by additional powers of $\partial ({\rm fields})/\Lambda$---become equally important and we no longer have a well-defined expansion. Note that the equations of motion following from the Lagrangian~\ref{simpleghostlag} are fourth order. To see explicitly that this implies the system secretly has a ghost, we introduce an auxiliary field, $\chi$, as follows~\cite{Creminelli:2005qk}
\be
{\cal L} = -\frac{1}{2}(\partial\psi)^2+\chi\square\psi-\frac{\Lambda^2}{2}\chi^2-V(\psi)~.
\label{auxchifield}
\ee
Now, the $\chi$ equation of motion is $\chi = \square\psi/\Lambda^2$, and substituting it back into the Lagrangian recovers~\ref{simpleghostlag}, confirming the classical equivalence between the two theories. In order to remove the kinetic mixing and diagonalize the Lagrangian, we may make the field redefinition $\psi = \phi-\chi$, in terms of which the Lagrangian becomes (after integration by parts)~\cite{Creminelli:2005qk}
\be
{\cal L} = -\frac{1}{2}(\partial\phi)^2+\frac{1}{2}(\partial\chi)^2-\frac{\Lambda^2}{2}\chi^2-V(\phi, \chi)~.
\ee
Thus, the Lagrangian~\ref{simpleghostlag} is equivalent to a theory of two scalar fields, one healthy and one ghostly. It is important to note that the presence of a ghost does not necessarily spell doom for a theory; as long as the mass of the ghostly mode lies above the cutoff of the theory, we can interpret the presence of the ghost as an artifact of truncating the EFT expansion at finite order. We see that this is the case here, the mass of the ghost lies at the cutoff of the theory, so it may be possible to consistently ignore the ghost at energies far below the cutoff and appeal to unknown UV physics to cure the ghost as we approach the cutoff. 

To see how this can work, instead of introducing the auxiliary field $\chi$ as in~\ref{auxchifield}, consider the Lagrangian~\cite{Creminelli:2005qk}
\be
{\cal L} = -\frac{1}{2}(\partial\psi)^2+\chi\square\psi-(\partial\chi)^2-\frac{\Lambda^2}{2}\chi^2-V(\psi)~.
\label{betterbehavedchilag}
\ee
The equation of motion for $\chi$ is
\be
\chi = \frac{\square\psi}{\Lambda^2\left(1-\frac{\square}{\Lambda^2}\right)}~;
\ee
substituting this back in recovers~\ref{simpleghostlag}, up to extra terms suppressed by powers of $\partial^2 ({\rm fields})/\Lambda^2$, which I have dropped in the EFT expansion anyway. Diagonalizing~\ref{betterbehavedchilag}, then reveals two healthy scalars
\be
{\cal L} = -\frac{1}{2}(\partial\phi)^2-\frac{1}{2}(\partial\chi)^2-\frac{\Lambda^2}{2}\chi^2-V(\phi,\chi)~.
\ee

What {\it is} dangerous is a theory possessing a ghost within the regime of validity of the effective theory. If this is true, the theory loses its predictive power. 

\subsection{Gradient instabilities}

Another pathology which often plagues effective field theories is the presence of {\it gradient} instabilities. Much in the way that a ghost instability is related to wrong sign temporal derivatives, gradient instabilities are due to wrong sign spatial gradients. To see why this is worrisome, consider the simplest (obviously non-Lorentz-invariant) example: a free scalar field with wrong sign spatial gradients
\be
{\cal L} = \frac{1}{2}\dot\phi^2+\frac{1}{2}(\vec\nabla\phi)^2~.
\label{gradinstablag}
\ee
The solutions to the equation of motion following from this Lagrangian are (in Fourier space)
\be
\phi_k(t) \sim e^{\pm kt}~,
\ee
where $k \equiv \sqrt{{\vec k}^2}$. Note that the growing mode solution, $\phi\sim e^{kt}$, grows without bound, signaling an instability in the theory on a timescale
\be
\tau_{\rm inst.} \sim k^{-1}~.
\ee
Therefore, the highest-energy modes contribute most to the instability. This means that the theory does not make sense, even thought of as an effective theory. In general, in a theory with a gradient instability, and a cutoff $\Lambda$, the effective theory cannot consistently describe any energy regime. For modes with $k \ll \Lambda$, the characteristic timescale is $t_k \gg \tau_{\rm inst.}$, and they will be sensitive to the instability in the theory, whereas modes with $k\gg \Lambda$ are beyond the regime of validity of our EFT. The conclusion is that an effective theory with a gradient instability is non-predictive.

\subsection{Tachyonic instabilities}

Another instability that can appear is the presence of a {\it tachyon}. Most simply, a tachyonic instability appears as a field with a negative mass squared. Unlike the other instabilities, the presence of a tachyon does not indicate any particular pathology in the definition of the theory, but rather is a signal that we are not perturbing about the true vacuum of the theory. For example, this is precisely what happens in the Higgs mechanism---in the Lagrangian, the Higgs field appears as a tachyon, but of course everything is well defined.

Again, this is most easily illustrated with a toy example: a scalar field with a negative mass term
\be
{\cal L} = -\frac{1}{2}(\partial\phi)^2+\frac{m^2}{2}\phi^2 \ .
\ee
In the long-wavelength ($k\to0$) limit, the solution for the field $\phi$ is
\be
\phi(t) \sim t^{\pm mt}~.
\ee
Again, the growing-mode solution indicates an instability. However, unlike last time, the timescale for this instability is independent of $k$ and is given by the inverse mass of the field
\be
\tau_{\rm inst.} \sim m^{-1}~.
\ee
Therefore, modes for which $k\gg m^{-1}$, will be insensitive to the fact that the system is unstable. This type of thinking is familiar from cosmology---if we go to high momenta, the modes evolve as though they are on Minkowski space and are insensitive to the cosmological evolution. 

This analysis generalizes to theories with a cutoff, $\Lambda$, in which there exists a regime $m \ll k \ll\Lambda$, where the effective field theory is perfectly well defined, provided that there is a hierarchy between the mass, $m$, and the cutoff of the theory.

\subsection{Analyticity, locality and superluminality}
\label{superlumapp}

So far, the pathologies we have discussed manifest themselves in the effective description (for example, ghosts are visible in the low energy effective theory). However, there are also apparent illnesses of an effective theory which are of a more subtle nature, and indicate that a theory---completely well-defined in the IR---may secretly not admit a standard UV completion. Whether or not such pathologies are fatal then depends on whether one can make sense of the relevant theories in the UV while abandoning one or more of the usual requirements, such as Lorentz-invariance. 

By far the most common sickness of this type is the presence of {\it superluminality} in a low-energy effective field theory. To understand why this is a problem, we note that a crucial ingredient of a Lorentz-invariant quantum field theory is {\it microcausality}. This property states that the commutator of two local operators vanishes for spacelike separated points as an operator statement~\cite{Peskin:1995ev}
\be
\left[{\cal O}_1(x), {\cal O}_2(y)\right] = 0~;~~~~~{\rm when}~~~~~(x-y)^2 > 0~.
\label{microcausalitystatement}
\ee
The relation to causality is fairly clear; if two operators are evaluated at points outside each others' lightcones, they should not have an effect on each other. Indeed, in~\cite{Dubovsky:2007ac}, it was shown that~\ref{microcausalitystatement} can be seen as a consequence of the causal structure of the theory, and hence holds for an arbitrary curved space, as long as the relevant fields have a well-defined Cauchy problem. Therefore, we immediately see that there is an apparent tension between superluminality and causality: in a theory with superluminal propagation, operators outside the light cone do not necessarily commute, indicating that the theory is secretly acausal or non-local.

In reality, theories which admit superluminality can be perfectly causal, but just on a widened light-cone. Consider a non-renormalizible higher derivative theory of a scalar
\be
{\cal L} = -\frac{1}{2}(\partial\phi)^2+\frac{1}{\Lambda^3}\partial^2\phi(\partial\phi)^2+\frac{1}{\Lambda^4}(\partial\phi)^4+\cdots
\label{nonlocallightcone}
\ee
expanded about some background $\phi = \bar\phi+\vp$. The causal structure is set by an {\it effective} metric~\cite{Wald:1984rg, Bruneton:2006gf,Dubovsky:2007ac,Babichev:2007dw}
\be
{\cal L} = -\frac{1}{2}G^{\mu\nu}(x,\bar\phi, \partial\bar\phi, \partial^2\bar\phi, \ldots)\partial_\mu\vp\partial_\nu\vp + \cdots \ .
\ee
Now, provided that $G^{\mu\nu}$ is globally hyperbolic, this theory will be perfectly causal, but in general it may have directions in which the $\vp$ perturbations propagate outside the lightcone used to define the theory~\ref{nonlocallightcone}. On the face of it, this might not seem to be much of a worry, but it is vaguely unsettling that the original Lorentz-invariant Lagrangian secretly may not be. Viewing the theory~\ref{nonlocallightcone} as a low-energy effective field theory, it must have a UV completion at some high energy scale. However, since this theory admits superluminality, it cannot be UV completed by a quantum field theory that is Lorentz-invariant with respect to the metric $\eta_{\mu\nu}$ used to define~\ref{nonlocallightcone}.

In many cases this heuristic reasoning can be made precise by exploiting the close relationship between Lorentz invariance and {\it S-matrix analyticity}. In a Lorentz-invariant quantum field theory, the S-matrix is an analytic function of the external momenta, except for branch cuts which correspond to the production of intermediate states and the presence of poles, which correspond to physical particles or bound states. From S-matrix analyticity, we can derive {\it dispersion relations} to establish the positivity of various scattering amplitudes. 

This can be clearly seen by considering the $2\to2$ scattering amplitude in a Lorentz-invariant theory. This amplitude intuitively should have something to do with superluminality, since propagation in the effective metric~\ref{nonlocallightcone} can be thought of as a sequence of scattering processes with a background field~\cite{Adams:2006sv}.
\begin{figure}
\centering
\includegraphics[width=2.2in]{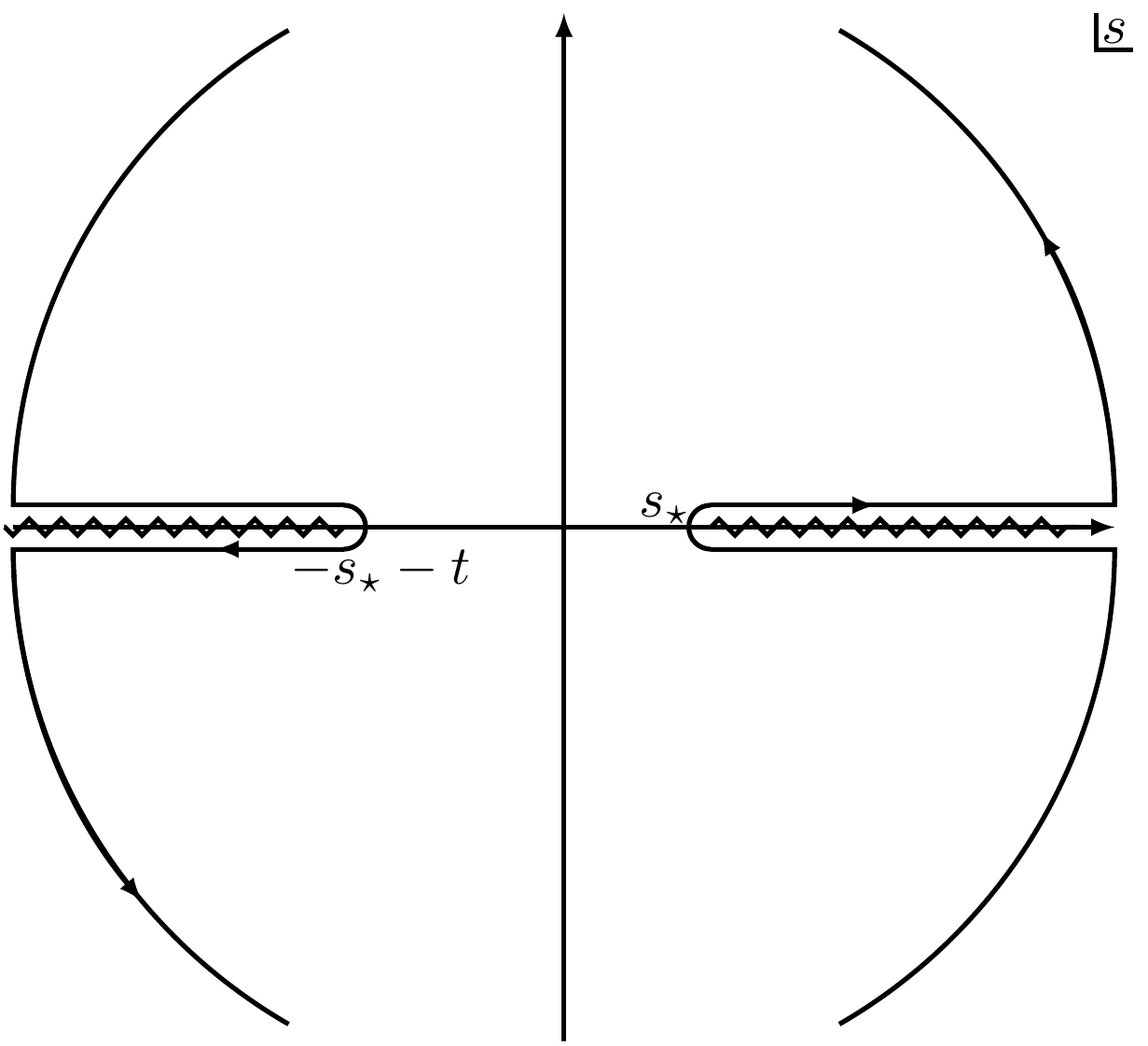}
\caption{\small Integration contour in the complex $s$-plane used to derive the dispersion relation. Integration along this contour picks up the discontinuity across the cuts, which correspond to above-threshold particle production.}
\label{dispersionint}
\end{figure}
Consider the $4$-point amplitude ${\cal A}(s, t)$, which is analytic in the $s$-plane, except for a pole at $s=0$ and along cuts on the real axis above some threshold value $| s_\star |< \infty$. Here $s$ and $t$ are the usual Mandelstam variables, and the parameter $s_\star$ is the energy at which we would expect to start pair producing particles in a scattering process (in a massive theory, this is the energy scale corresponding to the mass of the constituent particles). Considering a closed curve, $\gamma$, around $s=0$ of radius $r < s_\star$ and using Cauchy's integral formula, we find
\be
\left.\frac{\partial^2}{\partial s^2}{\cal A}(s, t)\right |_{s=0} = \frac{1}{i\pi}\oint_\gamma\rd s\frac{{\cal A}(s, t)}{s^3}~.
\ee
Now, we deform the contour into a double-keyhole contour and integrate along the cuts as in Figure~\ref{dispersionint}. On the positive real axis, we obtain the discontinuity along the cut, which is the imaginary part of the analytic function ${\cal A}$. The integral along the cut on the negative real axis can be related to the integral on the positive side via crossing symmetry, $s\to -t-s$, to obtain
\be
\frac{\partial^2}{\partial s^2}{\cal A}(s, t)\Big |_{s=0}  = \frac{2}{\pi}\int_{s_\star}^\infty\rd s\left(\frac{1}{s^3}+\frac{1}{(s+t)^3}\right){\rm Im}{\cal A}(s, t)~.
\label{dispersionrelation}
\ee
In order for this dispersion relation to make sense, we must make sure that the point at infinity gives no contribution to the integral. This contribution will vanish as long as the amplitude is bounded by $s^2$ as $s\to\infty$. That this is true in theories with a mass gap follows from the Froissart bound~\cite{Froissart:1961ux, Martin:1962rt, Adams:2006sv}. While many theories of interest do not have a mass gap (for example the galileon); there exist some arguments that amplitudes should be bounded similarly at infinity in these situations~\cite{Adams:2006sv}, although these results are not completely established. If we now look at the forward limit $(t\to0)$ of the expression~\ref{dispersionrelation}, the optical theorem tells us ${\rm Im}{\cal A}(s, 0) \geq 0$, which establishes the inequality
\be
\frac{\partial^2}{\partial s^2}{\cal A}(s, 0)\Big |_{s=0}  = \frac{4}{\pi}\int_{s_\star}^\infty\rd s\frac{{\rm Im}{\cal A}(s, 0)}{s^3} \geq 0~.
\label{dispersionsumrule}
\ee
Therefore, in the forward limit, the $2\to2$ amplitude must display a {\it positive} $s^2$ contribution.
This inequality holds in {\it any} Lorentz-invariant theory described by an S-matrix. This includes both local quantum field theory and perturbative string theories~\cite{Adams:2006sv}, making it a very powerful probe. Violation of this dispersion relation indicates a violation of Lorentz invariance in the theory.

\begin{figure}
\centering
\includegraphics[width=3.5in]{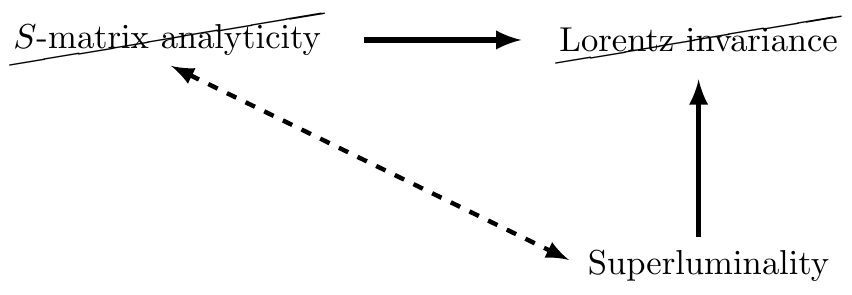}
\caption{\small Relationship between analyticity, locality and Lorentz invariance. Either non-analyticity of the $S$-matrix {\it or} superluminal propagation indicates a violation of Lorentz invariance. However, the implication between the two is less well understood.}
\end{figure}

We have seen that both superluminality and a violation of S-matrix analyticity indicate a violation of Lorentz invariance in a low energy effective theory. We might be tempted to posit that the relationship between these two things is tight, {\it i.e.}, that superluminality always implies that the theory will violate the sum rule~\ref{dispersionsumrule}, but unfortunately the relationship is somewhat more subtle. For the simplest theories, the relationship is tight, but it is possible to construct theories that admit apparently superluminal signals, but which obey~\ref{dispersionsumrule}~\cite{Nicolis:2009qm, Hinterbichler:2012yn, deRham:2013hsa}. Here we have only focused on the simplest dispersion relation, but it is possible to extract more intricate ones using similar arguments. It is generally expected that theories which admit superluminality should violate some dispersion relations, as they follow directly from Lorentz invariance and analyticity and we know that superluminal theories are secretly not Lorentz-invariant, but there is no known single dispersion relation that is always violated by a superluminal theory.
In any case, in order for a low-energy effective theory to be UV-completable by a local, Lorentz invariant QFT or perturbative string theory, it must satisfy the dispersion relation~\ref{dispersionsumrule} (along with an infinite number of other ones) and it must not have superluminal signals in the effective theory.

Even the issue of whether a theory actually exhibits superluminality is somewhat subtle, as is emphasized in~\cite{deRham:2014lqa}, and the fact that a theory exhibits a superluminal dispersion relation at tree level in the Lagrangian is not sufficient to conclude that the theory is acausal. One concrete example of this type of phenomenon occurs in the context of galileon duality~\cite{deRham:2013hsa}.

\section{Example: The Galileons}
Galileons provide a topical and relatively simple playground in which to explore many of these topics.
An easy starting point from which to motivate Galileons is the Dvali-Gabadadze-Porrati (DGP) model~\cite{Dvali:2000hr}, in which a $3+1$-dimensional brane is embedded in a $5$d bulk, with action
\be
S=\frac{M_5^3}{2}\int d^5X\, \sqrt{-G}\,\  R[G] + \frac{M_4^2}{2}\int d^4x\, \sqrt{-g}\, \ R[g] \ .
\label{DGPaction}
\ee
In this model gravity is modified on large distances, and there exists a branch of cosmological solutions that exhibit self-acceleration at late times, although this branch contains a ghost. There also exists a ghost-free ``normal" branch, for which a 4d effective action can be derived which, in a particular ``decoupling" limit, reduces to a theory of a single scalar $\phi$, with a cubic self-interaction term $\sim (\partial\phi)^2\Box \phi$.  This term yields second order field equations and is invariant (up to a total derivative) under the {\it Galilean} transformations
\be 
\label{Galileoninvarianceold}
\phi (x) \rightarrow \phi(x) + c + b_\mu x^\mu \ ,
\ee
with $c$ and $b_{\mu}$ constants.  

It is natural to ask what other terms are allowed in a four dimensional field theory with this same symmetry~\cite{Nicolis:2008in}. Referring to the field of such a theory as the {\it Galileon}, it turns out that, in addition to an infinite number of terms in which two derivatives act on every field (trivially invariant under the galilean symmetry) there are a finite number of terms that have fewer numbers of derivatives per field. These possess second order equations of motion, and there can exist regimes in which these special terms are important, and the infinity of other possible terms within the effective field theory are not. Along with a non-renormalization theorem for Galileons~\cite{Luty:2003vm,Hinterbichler:2010xn,Burrage:2010cu}, this means that we are able, in some circumstances, to understand non-linear effects that are quantum mechanically exact.

For $n\geq 1$, the $(n+1)$-th order Galileon Lagrangian (where ``order" refers to the number of copies of the scalar field $\phi$ that appear in the term) is
\be
\label{Galileon2} 
{\cal L}_{n+1}=n\eta^{\mu_1\nu_1\mu_2\nu_2\cdots\mu_n\nu_n}\left( \partial_{\mu_1}\phi\partial_{\nu_1}\phi\partial_{\mu_2}\partial_{\nu_2}\phi\cdots\partial_{\mu_n}\partial_{\nu_n}\phi\right) \ ,
\ee 
where 
\be
\label{tensor} 
\eta^{\mu_1\nu_1\mu_2\nu_2\cdots\mu_n\nu_n}\equiv{1\over n!}\sum_p\left(-1\right)^{p}\eta^{\mu_1p(\nu_1)}\eta^{\mu_2p(\nu_2)}\cdots\eta^{\mu_np(\nu_n)} 
\ee 
and the sum is over all permutations of the $\nu$ indices, with $(-1)^p$ the sign of the permutation.  The tensor~(\ref{tensor}) is antisymmetric in the $\mu$ indices, antisymmetric in the $\nu$ indices, and symmetric under interchange of any $\mu,\nu$ pair with any other.  These Lagrangians are unique up to total derivatives and overall constants.   Because of the antisymmetry requirement on $\eta$, only the first $n$ of these Galileons are non-trivial in $n$-dimensions.  In addition, the tadpole term, $\phi$, is Galilean invariant, and we therefore include it as the first-order Galileon.  These Galileon actions can be generalized to the multi-field case, where there is a multiplet $\phi^I$ of fields~\cite{Deffayet:2010zh,Padilla:2010de,Padilla:2010ir,Hinterbichler:2010xn}, while retaining many of the same desirable properties, and can be constructed as Wess-Zumino terms of the nonlilnearly realized galileon symmetry~\cite{Goon:2012dy}.

It should be remembered that all these theories are not renormalizable, and should be considered as effective field theories with some cutoff $\Lambda$, above which some UV completion is required.   However, as I have mentioned, there can exist regimes in which the above terms dominate, since the symmetries forbid any renormalizable terms, and other terms will have more derivatives.   One important consequence of these properties is that galileon theories exhibit a particular type of screening mechanism - the Vainshtein mechanism~\cite{Vainshtein:1972sx} - that allows them to mediate long range forces, competing with gravity at cosmic scales, while having negligible physical effects at the scales of local tests of gravity. This is the key to their use in modified gravity applications (see~\cite{Joyce:2014kja} for a recent review of these mechanisms and their effects and tests).

\subsubsection{Solution around spherically-symmetric source}

This is perhaps best illustrated by considering the emergence of Vainshtein screening near a static point source of mass $M$, so that $T^\mu_{\;\mu} =-M\delta^{(3)}(\vec{x})$.  Assuming the field profile is static and  spherically-symmetric, ($\phi = \phi(r)$), the equation of motion reduces to
\be
\vec{\nabla} \cdot \left(6\vec\nabla\phi + \hat{r} \frac{4}{\Lambda^3}\frac{(\vec\nabla\phi)^2}{r}\right) = \frac{gM}{M_{\rm Pl}} \delta^{(3)}(\vec{x})\,.
\ee
Integrating over a sphere centered at the origin, we obtain
\be
6\phi' + \frac{4}{\Lambda^3}\frac{\phi'^2}{r} = \frac{gM}{4\pi r^2 M_{\rm Pl}}\, .
\ee
This equation is now algebraic in $\phi'$, and so admits a solution by radicals. Focusing on the branch for which $\phi'\rightarrow 0$ at spatial infinity,
\be
\phi'(r) = \frac{3\Lambda^3  r}{4} \left( -1+ \sqrt{1 + \frac{1}{9\pi}\left(\frac{r_{\rm V}}{r}\right)^3 }\right)\, ,
\label{galfieldprofile}
\ee
where we have introduced the {\it Vainshtein radius}
\be
r_{\rm V} \equiv \frac{1}{\Lambda } \left(\frac{gM}{M_{\rm Pl}}\right)^{1/3}\ .
\ee
\begin{figure}[tb]
\begin{center}
\includegraphics[width=6in]{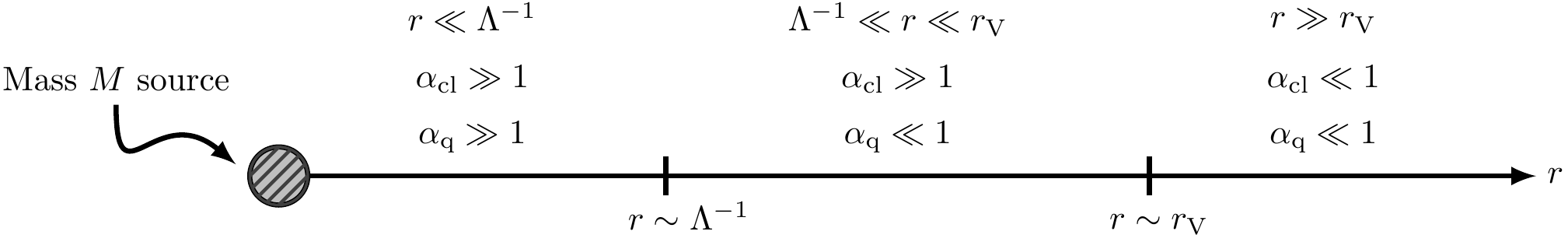}
\caption{\small Various regimes of the galileon theory around a spherically symmetric source. Beyond the Vainshtein radius, $r_{\rm V} \sim (M/\Lambda^3 M_{\rm Pl})^{1/3}$, the field mediates a long range force and both the classical, $\alpha_{\rm cl} \sim (r_{\rm V}/r)^3$, and quantum, $\alpha_{\rm q}\sim (r\Lambda)^{-2}$, non-linearity parameters are small. Very close to the source---$r\ll \Lambda^{-1}$, the inverse cutoff---both $\alpha_{\rm cl}$ and $\alpha_{\rm q}$ are large and the theory is not predictive. However, there is an intermediate regime, $\Lambda\ll r\ll r_{\rm V}$, where classical non-linearities are important ($\alpha_{\rm cl}\gg1$), but quantum effects can consistently be neglected ($\alpha_{\rm q}\ll 1$).
}
  \label{vainshteinregime}
\end{center}
\end{figure}
This nontrivial radial profile is crucial to the operation of Vainshtein screening. We consider two regimes:

\begin{itemize}

\item $r \gg r_{\rm V}$: Far away from the source, the solution is approximately a $1/r^2$ profile,
\be
\phi'(r\gg r_{\rm V}) \simeq \frac{g}{3} \cdot \frac{M}{8\pi M_{\rm Pl} r^2}\,.
\ee
In this regime, the galileon force relative to gravity is given by
\be
\left. \frac{F_{\phi}}{F_{\rm gravity}}\right\vert_{r\gg r_{\rm V}}\simeq \frac{g^2}{3}\, ,
\ee
and both the classical and quantum expansion parameters are small (we assume $M \gg M_{\rm Pl}$, so that $\Lambda^{-1} \ll r_{\rm V}$):
\be
\alpha_{\rm cl} \sim \left(\frac{r_{\rm V}}{r}\right)^3 \ll 1\,~~~~~~~~~~~~~\alpha_{\rm q} \sim \frac{1}{(r\Lambda)^2} \ll 1\,.
\ee
This tells us that both classical non-linearities and quantum corrections are unimportant.

\item $r \ll r_{\rm V}$: Near the source,~\ref{galfieldprofile} reduces to
\be
\phi'(r\ll r_{\rm V})  \simeq \frac{\Lambda^3r_{\rm V}}{2} \sqrt{\frac{r_{\rm V}}{r}}\sim \frac{1}{\sqrt{r}} \,.
\label{Er<<rV}
\ee
The force due to the galileon relative to that of gravity is now given by
\be
\left. \frac{F_{\phi}}{F_{\rm gravity}}\right\vert_{r\ll r_{\rm V}} \sim \left(\frac{r}{r_{\rm V}}\right)^{3/2}\ll 1\,,
\label{galforcesup}
\ee
so the scalar force is strongly suppressed at distances much less than the Vainshtein radius.
In this regime, the classical non-linearity parameter is very large (as it must be, this is the source of the screening),
\be
\alpha_{\rm cl} \sim  \left(\frac{r_{\rm V}}{r}\right)^{3/2} \gg 1\,.
\ee
But notice that the quantum parameter is {\it not}; it takes the same form as before:
\be
\alpha_{\rm q} \sim \frac{1}{(r\Lambda)^2} \,.
\ee
At distances $r \gg \Lambda^{-1}$, this is small and quantum corrections are under control, meaning that the classical solution can be trusted. Of course, sufficiently close to the source,
$r \ll \Lambda^{-1}$, the quantum parameter becomes ${\cal O}(1)$, radiative corrections become important, and the effective field theory breaks down. (In fact, this statement is too conservative---we will see shortly that perturbations
acquire a large kinetic term scaling as $\sim (r_{\rm V}/r)^{3/2}$. Upon canonical normalization, this translates to a higher strong coupling scale. Even ignoring this fact, the scale $\Lambda$ is only the {\it strong-coupling} scale, it may be possible to re-sum the quantum corrections into a predictive theory.)

\end{itemize}
Therefore, as advertised, we see that there exists a regime,  $ \Lambda^{-1} \ll r \ll r_{\rm V}$, where classical non-linearities are important while quantum effects remain small.

\subsubsection{Perturbations around the spherically-symmetric background}

Above we have considered the background field profile around a massive source, but the Vainshtein mechanism can be further understood by considering perturbations about this solution. We can consider linearized perturbations by expanding~\ref{L3} as $\varphi = \phi - \bar{\phi}$, $T_{\mu\nu} = T_{\mu\nu}+\delta T_{\mu\nu}$ gives
\be
{\cal L}_{\varphi} = \left[ 3+  \frac{2}{\Lambda^3}\left(\bar\phi'' + \frac{2\bar\phi'}{r} \right)\right] \left(\dot{\varphi}^2 - (\partial_\Omega \varphi)^2\right) - \left[3 + \frac{4}{\Lambda^3}\frac{\bar\phi'}{r} \right] (\partial_r\varphi)^2 - \frac{1}{\Lambda^3}\Box\varphi(\partial\varphi)^2 + \frac{g}{M_{\rm Pl}} \varphi \delta T^\mu_{\;\mu} \ ,
\ee
where $\partial_\Omega$ denotes the usual angular derivatives. If we then look deep inside the Vainshtein radius ($r \ll r_{\rm V}$), by substituting the expression~\ref{Er<<rV} for $E$, we obtain
\be
{\cal L}_{\varphi}\sim \left(\frac{r_{\rm V}}{r}\right)^{3/2} \left(\dot{\varphi}^2 - (\partial_\Omega \varphi)^2 - \frac{4}{3} (\partial_r\varphi)^2 \right) - \frac{1}{\Lambda^3}\Box\varphi(\partial\varphi)^2 + \frac{g}{M_{\rm Pl}} \varphi \delta T^\mu_{\;\mu}.
\label{L3varphi}
\ee
The key thing to notice in this expression is that an enhancement factor of  $(r_{\rm V}/r)^{3/2} \gg 1$ multiplies the kinetic term, telling us that perturbations acquire a large inertia near a massive source.
Said differently, performing the canonical normalization $\varphi_{\rm c} \equiv \left(\frac{r_{\rm V}}{r}\right)^{3/4} \varphi$, the effective coupling to matter is reduced to
\be
g_{\rm eff} \sim \left(\frac{r}{r_{\rm V}}\right)^{3/4} g \ll g\,;
\ee
this indicates that galileon perturbations decouple from matter. Further, the strong coupling scale $\Lambda$ is dressed to a higher scale
\be
\Lambda^{\rm eff} \sim \left(\frac{r_{\rm V}}{r}\right)^{3/4}  \Lambda~\gg \Lambda\,,
\ee
which leads the perturbations to have weaker self-interactions.  

\begin{figure}
\centering
\includegraphics[width=3.3in]{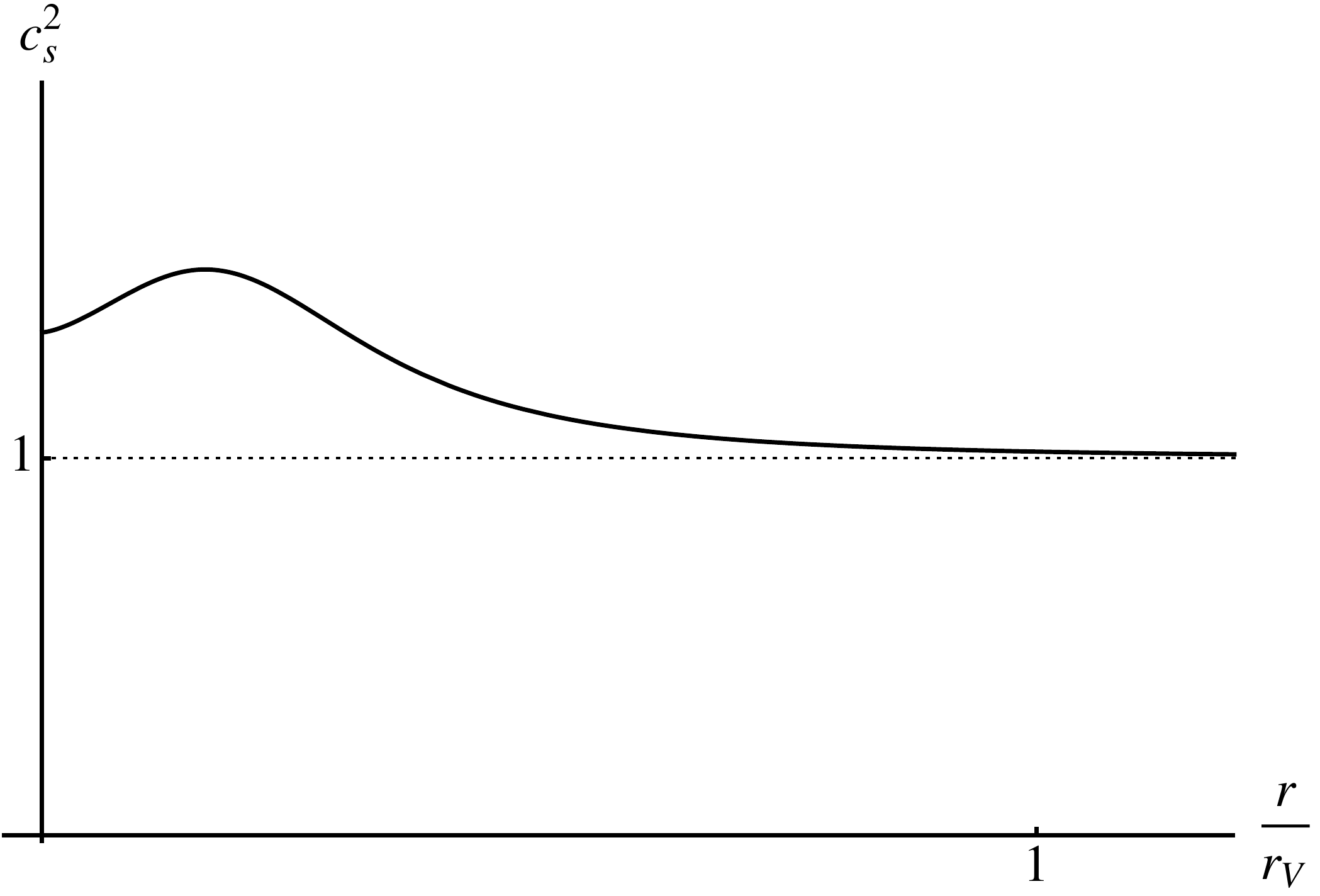}
\caption{\label{DGPsuperluminal}\small Plot of speed of radial fluctuations versus distance from a spherically symmetric source, in units of $r_{\rm V}$ for the cubic galileon.}
\end{figure}

Another thing to notice from~\ref{L3varphi} is that the radial speed of propagation is {\it superluminal}:
\be
c_{\rm s}^{\rm radial} = \sqrt{\frac{4}{3}}\,.
\ee 
This superluminality is a generic feature of galileons---galileon interactions are derivative interactions, so a galileon background (even one which is arbitrarily weak) deforms the
light-cone for perturbations in such a way that there is always a direction in which the speed of propagation is superluminal~\cite{Nicolis:2009qm}. With superluminality
comes the risk of the standard paradoxes due to closed time-like curves (CTC).
The situation is actually not as bad as one might fear. To start with, galileons by themselves are completely fine---the effective light-cone for the metric governing perturbations, albeit wider than the Minkowskian light-cone, admits a well-defined causal structure. In other words, galileons by themselves cannot generate CTCs. On the other hand, CTCs become possible when considering galileons coupled to (Lorentz-invariant) matter. However, it was conjectured in~\cite{Burrage:2011cr} that galileons are protected from the formation of CTCs by a Chronology Protection Criterion, analogous to that of GR, so that if one starts with healthy initial conditions and tries to construct a CTC, the galileon effective field theory will break down before it can form. However, as we have discussed, the existence of superluminal propagation around certain backgrounds signals that the UV completion of galileons, if one exists, is not a local (Lorentz-invariant) quantum field theory but something more exotic (or more interesting, depending on one's point of view). Recent arguments suggest that this apparent superluminality might be an artifact of trusting a tree-level computation in a regime where it is unreliable~\cite{deRham:2013hsa, deRham:2014lqa}, so the severity of this peculiarity is far from settled at present.
A related tension is that the galileon terms lead to scattering amplitudes which do not obey dispersion relations obtained from the arguments about S-matrix analyticity we have discussed earlier. On the other hand, this argument relies on the existence of an S-matrix for galileons, which has been questioned~\cite{Berezhiani:2013dca,Berezhiani:2013dw}.

\section{Conclusions}
In this brief summary of comments made at the Dark Side of the Universe conference, I have described theoretical issues that arise in a number of models proposed to address the accelerated expansion of the universe. While these issues must be kept in mind for any theory, in those that exhibit screening effects, which are often required to make accelerating models viable, particularly interesting constraints can arise. As an example, I have focused on the galileon model, which exhibits the Vainshtein effect, necessary to avoid precise solar system tests of gravity, and have discussed how this screening works and how some of the theoretical constraints arise. 

The considerations described here are quite general and all or a subset of them must be evaded, or at the very least understood, in any putative theory of late-time cosmic acceleration, such as dark energy or modified gravity.

\acknowledgments
I would like to thank my hosts at the Yukawa Institute for Theoretical Physics for such an enjoyable conference, and several of my collaborators - Garrett Goon, Kurt Hinterbichler, Bhuvnesh Jain, Austin Joyce and Justin Khoury - for their permission to reproduce large portions of our joint work essentially verbatim in this proceedings. This work was supported in part by US Department of Energy grant DE-SC0013528, and by NASA ATP grant NNX11AI95G.


\begin{thebibliography}{99}
\bibitem{Joyce:2014kja} 
  A.~Joyce, B.~Jain, J.~Khoury and M.~Trodden,
  Phys.\ Rept.\  {\bf 568}, 1 (2015)
  [arXiv:1407.0059 [astro-ph.CO]].
  
\bibitem{Carroll:2003st} 
  S.~M.~Carroll, M.~Hoffman and M.~Trodden,
  Phys.\ Rev.\ D {\bf 68}, 023509 (2003)
  [astro-ph/0301273].
  
\bibitem{Cline:2003gs} 
  J.~M.~Cline, S.~Jeon and G.~D.~Moore,
  Phys.\ Rev.\ D {\bf 70}, 043543 (2004)
  [hep-ph/0311312].
  
\bibitem{Creminelli:2005qk} 
  P.~Creminelli, A.~Nicolis, M.~Papucci and E.~Trincherini,
  JHEP {\bf 0509}, 003 (2005)
  [hep-th/0505147].
  
\bibitem{Peskin:1995ev} 
  M.~E.~Peskin and D.~V.~Schroeder,
  Reading, USA: Addison-Wesley (1995) 842 p
  
\bibitem{Dubovsky:2007ac} 
  S.~Dubovsky, A.~Nicolis, E.~Trincherini and G.~Villadoro,
Phys.\ Rev.\ D {\bf 77}, 084016 (2008)
[arXiv:0709.1483 [hep-th]].
  
\bibitem{Wald:1984rg} 
  R.~M.~Wald,
Chicago, Usa: Univ. Pr. ( 1984) 491p

\bibitem{Bruneton:2006gf} 
  J.~P.~Bruneton,
Phys.\ Rev.\ D {\bf 75}, 085013 (2007)
[gr-qc/0607055].

\bibitem{Babichev:2007dw} 
  E.~Babichev, V.~Mukhanov and A.~Vikman,
JHEP {\bf 0802}, 101 (2008)
[arXiv:0708.0561 [hep-th]].
  
\bibitem{Adams:2006sv} 
  A.~Adams, N.~Arkani-Hamed, S.~Dubovsky, A.~Nicolis and R.~Rattazzi,
 JHEP {\bf 0610}, 014 (2006)
  [hep-th/0602178].
  
\bibitem{Froissart:1961ux} 
  M.~Froissart,
Phys.\ Rev.\  {\bf 123}, 1053 (1961).

\bibitem{Martin:1962rt} 
  A.~Martin,
Phys.\ Rev.\  {\bf 129}, 1432 (1963).

\bibitem{Nicolis:2009qm} 
  A.~Nicolis, R.~Rattazzi and E.~Trincherini,
JHEP {\bf 1005}, 095 (2010)
Erratum: [JHEP {\bf 1111}, 128 (2011)]
[arXiv:0912.4258 [hep-th]].

\bibitem{Hinterbichler:2012yn} 
  K.~Hinterbichler, A.~Joyce, J.~Khoury and G.~E.~J.~Miller,
Phys.\ Rev.\ Lett.\  {\bf 110}, no. 24, 241303 (2013)
[arXiv:1212.3607 [hep-th]].

\bibitem{deRham:2013hsa} 
  C.~de Rham, M.~Fasiello and A.~J.~Tolley,
  Phys.\ Lett.\ B {\bf 733}, 46 (2014)
  [arXiv:1308.2702 [hep-th]].

\bibitem{deRham:2014lqa} 
  C.~De Rham, L.~Keltner and A.~J.~Tolley,
  Phys.\ Rev.\ D {\bf 90}, no. 2, 024050 (2014)
  [arXiv:1403.3690 [hep-th]].
  
\bibitem{Dvali:2000hr} 
  G.~R.~Dvali, G.~Gabadadze and M.~Porrati,
  Phys.\ Lett.\ B {\bf 485}, 208 (2000)
  [hep-th/0005016].
  
\bibitem{Nicolis:2008in} 
  A.~Nicolis, R.~Rattazzi and E.~Trincherini,
  Phys.\ Rev.\ D {\bf 79}, 064036 (2009)
  [arXiv:0811.2197 [hep-th]].
  
\bibitem{Luty:2003vm} 
  M.~A.~Luty, M.~Porrati and R.~Rattazzi,
JHEP {\bf 0309}, 029 (2003)
[hep-th/0303116].

\bibitem{Hinterbichler:2010xn} 
  K.~Hinterbichler, M.~Trodden and D.~Wesley,
Phys.\ Rev.\ D {\bf 82}, 124018 (2010)
[arXiv:1008.1305 [hep-th]].

\bibitem{Burrage:2010cu} 
  C.~Burrage, C.~de Rham, D.~Seery and A.~J.~Tolley,
JCAP {\bf 1101}, 014 (2011)
[arXiv:1009.2497 [hep-th]].

\bibitem{Deffayet:2010zh} 
  C.~Deffayet, S.~Deser and G.~Esposito-Farese,
Phys.\ Rev.\ D {\bf 82}, 061501 (2010)
[arXiv:1007.5278 [gr-qc]].

\bibitem{Padilla:2010de} 
  A.~Padilla, P.~M.~Saffin and S.~Y.~Zhou,
JHEP {\bf 1012}, 031 (2010)
[arXiv:1007.5424 [hep-th]].

\bibitem{Padilla:2010ir} 
  A.~Padilla, P.~M.~Saffin and S.~Y.~Zhou,
Phys.\ Rev.\ D {\bf 83}, 045009 (2011)
[arXiv:1008.0745 [hep-th]].

\bibitem{Goon:2012dy} 
  G.~Goon, K.~Hinterbichler, A.~Joyce and M.~Trodden,
JHEP {\bf 1206}, 004 (2012)
[arXiv:1203.3191 [hep-th]].

\bibitem{Vainshtein:1972sx} 
  A.~I.~Vainshtein,
Phys.\ Lett.\ B {\bf 39}, 393 (1972).

\bibitem{Burrage:2011cr} 
  C.~Burrage, C.~de Rham, L.~Heisenberg and A.~J.~Tolley,
JCAP {\bf 1207}, 004 (2012)
[arXiv:1111.5549 [hep-th]].

\bibitem{Berezhiani:2013dca} 
  L.~Berezhiani, G.~Chkareuli, C.~de Rham, G.~Gabadadze and A.~J.~Tolley,
Class.\ Quant.\ Grav.\  {\bf 30}, 184003 (2013)
[arXiv:1305.0271 [hep-th]].

\bibitem{Berezhiani:2013dw} 
  L.~Berezhiani, G.~Chkareuli and G.~Gabadadze,
Phys.\ Rev.\ D {\bf 88}, 124020 (2013)
[arXiv:1302.0549 [hep-th]].
  
  \end{thebibliography}
\end{document}